\documentclass[12pt]{iopart_mod}

\pdfoutput=1

\usepackage{iopams} 
\usepackage{mathtools}
\usepackage{amssymb,amsmath,amsfonts}

\usepackage[
      colorlinks=true,
      linkcolor=blue,
      urlcolor=blue,
      filecolor=black,
      citecolor=red,
      pdfstartview=FitV,
      pdftitle={},
        pdfauthor={Daniel Grumiller, Wout Merbis and Max Riegler},
        pdfsubject={},
        pdfkeywords={},
        pdfpagemode=None,
        bookmarksopen=true
      ]{hyperref}

\usepackage[all]{hypcap}

\DeclareFontFamily{OT1}{rsfs}{}
\DeclareFontShape{OT1}{rsfs}{m}{n}{ <-7> rsfs5 <7-10> rsfs7 <10->rsfs10}{} 
\DeclareMathAlphabet{\mycal}{OT1}{rsfs}{m}{n}

\newcommand{\be}[1]{ \begin{equation}\label{#1} }
\newcommand{\ee}{\end{equation}}
\newcommand{\bea}[1]{\begin{eqnarray}\label{#1} }
\newcommand{\eea}{\end{eqnarray}}

\newcommand{\eq}[2]{\begin{equation} #1 \label{#2} \end{equation}}

\DeclareMathOperator{\extdm}{d}
\newcommand{\extd}{\extdm \!}

\newcommand{\vp}{\varphi}
\newcommand{\cL}{\mathcal{L}}
\newcommand{\cM}{\mathcal{M}}
\newcommand{\cO}{\mathcal{O}}

\usepackage{xcolor}
\usepackage[makeroom]{cancel}

\definecolor{color1}{rgb}{0.22,0.45,0.70}

\begin{document}

\title{Most general flat space boundary conditions in three-dimensional Einstein gravity}

\author{Daniel Grumiller, Wout Merbis}
\address{Institute for Theoretical Physics, TU Wien, Wiedner Hauptstr.~8-10/136, A-1040 Vienna, Austria, Europe}
\ead{grumil@hep.itp.tuwien.ac.at, merbis@hep.itp.tuwien.ac.at}
\author{Max Riegler}
\address{Universit{\'e} libre de Bruxelles, Boulevard du Triomphe (Campus de la Plaine), 1050 Bruxelles,
Belgium, Europe}
\ead{max.riegler@ulb.ac.be}
\vspace{10pt}
\begin{indented}
\item[]March 2017
\end{indented}

\begin{abstract}
We consider the most general asymptotically flat boundary conditions in three-dimensional Einstein gravity in the sense that we allow for the maximal number of independent free functions in the metric, leading to six towers of boundary charges and six associated chemical potentials. We find as associated asymptotic symmetry algebra an $\mathfrak{isl}(2)_k$ current algebra. Restricting the charges and chemical potentials in various ways recovers previous cases, such as $\mathfrak{bms}_3$, Heisenberg or Detournay--Riegler, all of which can be obtained as contractions of corresponding AdS$_3$ constructions.  Finally, we show that a flat space contraction can induce an additional Carrollian contraction. As examples we provide two novel sets of boundary conditions for Carroll gravity.  
\end{abstract}

\vspace{2pc}

\submitto{\CQG BMS Focus Issue}
\setcounter{footnote}{0}

\section{Introduction}

Far away from gravitational sources flat (Minkowski) space seems a reasonable approximation as long as effects from a cosmological constant can be neglected. Intuitively, it makes sense to expect that special relativity is the limit of general relativity in the limit of weak gravitational fields. Nevertheless, this intuition is misguided, as famously shown by Bondi, van der Burg, Metzner and Sachs in the early 1960ies \cite{Bondi:1962, Sachs:1962}. Namely, the isometries of Minkowski space generate only the Poincar\'e algebra, while the asymptotic isometries of asymptotically flat spacetimes generate the infinite dimensional Bondi--Metzner--Sachs (BMS) algebra \cite{Bondi:1962, Sachs:1962}, which includes an infinite set of so-called ``supertranslations'' on top of the Lorentz algebra.

Thus, scattering processes for massless fields in asymptotically flat spacetimes are governed not just by the Poincar\'e algebra, but by the full BMS algebra (or perhaps an even larger algebra). One manifestation of this fact is the equivalence of Weinberg's soft graviton theorem \cite{Weinberg:1965nx} to supertranslation-invariance of the gravitational S-matrix \cite{Strominger:2013jfa, He:2014laa}.

BMS is not the end of the asymptotically flat story, though. Newman and Unti discovered slightly more general asymptotically flat boundary conditions \cite{Newman:1962cia}, which enhances BMS$_4$ by conformal rescalings of the metric on null infinity. Barnich and Troessaert also advocated more general boundary conditions that enhance not only the Poincar\'e-translations to supertranslations, but also the Lorentz-rotations to ``superrotations'' (local conformal transformations of the celestial sphere) \cite{Barnich:2009se, Barnich:2010eb}. Interestingly, in the Barnich--Troessaert setup Weinberg's soft graviton theorem generalizes and subleading terms therein are subject to superrotation-invariance \cite{Cachazo:2014fwa, Kapec:2014opa}. For a detailed account of BMS symmetries see for instance \cite{Oblak:2016eij} and references therein.

Is it then fair to ask: Is Barnich--Troessaert the end of the asymptotically flat story? Or are there more general/different asymptotically flat boundary conditions that potentially feature in statements about S-matrices for massless (soft) particles?

Indeed, there are indications that a further generalization of the superrotation group to arbitrary diffeomorphisms of the celestial 2-sphere is possible and useful \cite{Campiglia:2014yka, Campiglia:2015yka, Penna:2015gza}, but it is still not obvious if that further generalization would be the end of the story that was started by BMS in the 1960ies. Given these developments and open issues it is of interest to find the loosest consistent set of asymptotically flat boundary conditions and to recover previous cases (and possibly novel ones) as restrictions imposed on this loosest set. This motivates our present work. However, we are not going to address the four-dimensional case and instead focus on three spacetime dimensions.

While the original BMS construction was done only in four spacetime dimensions, their methods were applied and generalized to other dimensions as well. Of particular interest is the situation in three spacetime dimensions \cite{Ashtekar:1996cd} where the BMS algebra acquires a non-trivial central extension \cite{Barnich:2006av}. In the past five years centrally extended $\mathfrak{bms}_3$ symmetries were exploited for numerous aspects of flat space holography like microstate Cardyology \cite{Barnich:2012xq, Bagchi:2012xr}, (holographic) entanglement entropy \cite{Bagchi:2014iea,Basu:2015evh} or holographic calculation of stress tensor correlators \cite{Bagchi:2015wna}, see \cite{Riegler:2016hah, Bagchi:2016bcd} for reviews.

In the present work we find the most general asymptotically flat boundary conditions for three-dimensional Einstein gravity.
This paper is organized as follows. 
In section \ref{se:2} we review flat space Einstein gravity in the Chern--Simons formulation.
In section \ref{se:3} we present the most general flat space boundary conditions and derive the associated asymptotic symmetry algebra.
In section \ref{se:4} we translate results into the metric formulation.
In section \ref{se:5} we discuss various special cases, recovering all existing flat space boundary conditions.
In section \ref{se:ads} we consider limits from AdS and find novel boundary conditions for Carroll gravity.
In section \ref{se:6} we conclude with a discussion of various checks, generalizations and open issues.

\section{Flat space Einstein gravity in Chern--Simons formulation}\label{se:2}

The three-dimensional Einstein--Hilbert action \cite{Staruszkiewicz:1963zz, Deser:1984tn}
\eq{
I_{\textrm{\tiny EH}}[g] = \frac{1}{16\pi G}\,\int\extd^3x\sqrt{|g|}\,R
}{eq:bms1}
is classically equivalent to the Chern--Simons action \cite{Achucarro:1987vz, Witten:1988hc} up to boundary terms 
\eq{
I_{\textrm{\tiny CS}}[A] = \frac{k}{4\pi}\,\int\langle A\wedge\extd A + \tfrac23\,A\wedge A \wedge A\rangle\,.
}{eq:bms2}
The Chern--Simons level $k$ is related to the three-dimensional Newton constant $G$ by $k=1/(4G)$ and the Chern--Simons connection $A$ is an element of $\mathfrak{isl}(2)$. 

Explicitly, we decompose the connection 
\eq{
A = e^n M_n + \omega^n L_n
}{eq:bms3}
in terms of the $\mathfrak{isl}(2)$ generators $L_n, M_n$ ($n=\pm 1,0$), whose commutations relations read
\begin{align}
[L_n, L_m] & = (n-m)L_{m+n} & [L_m,M_n] & = (n-m)M_{m+n} & [M_n,M_m] & = 0\,.
\end{align}
Note that $\mathfrak{isl}(2)$ is the global part of (centrally extended) $\mathfrak{bms}_3$, in the same way as $\mathfrak{sl}(2)\oplus\mathfrak{sl}(2)$ is the global part of the two-dimensional conformal algebra. 

The $M_n$-part of the connection $A$ corresponds to the dreibein $e$ (the $L_n$-part corresponds to the dualized spin connection $\omega$). The map to the metric formulation is then determined by a ``twisted'' trace that picks up only the $M_n$-part of the connection \cite{Gary:2014ppa}
\begin{equation}
g_{\mu\nu}\extd x^{\mu} \extd x^{\nu} = \widetilde{\tr}\big(A A\big) = \kappa_{nm} e^n e^m = - 4 e^+ e^- + (e^0)^2\,.
\label{eq:ttr}
\end{equation}
The trace appearing in the action is the $\mathfrak{isl}(2)$ invariant bilinear form
\begin{equation}
\langle L_mM_n\rangle =\kappa_{nm} = \left( \begin{array}{ccc} 0 & 0 & -2 \\ 0 & 1 & 0 \\ -2 & 0 & 0\end{array} \right)
\label{eq:bi}
\end{equation}
with $\langle L_mL_n\rangle=\langle M_mM_n\rangle=0$ and contains the same matrix $\kappa_{nm}$ as the twisted trace.

Instead of imposing boundary conditions on the metric one can equivalently impose corresponding boundary conditions on the gauge connection $A$. This is precisely what we do in the next section, thereby generalizing the recent analysis in AdS$_3$ to flat space \cite{Grumiller:2016pqb}.

\section{Most general flat space boundary conditions}\label{se:3}

In this section we formulate the loosest set of flat space boundary conditions for three-dimensional Einstein gravity, following closely the similar construction in AdS$_3$ \cite{Grumiller:2016pqb}.
In section \ref{se:2.1} we state the boundary conditions. 
In section \ref{se:2.2} we derive all transformations that preserve the boundary conditions.
In section \ref{se:2.3} we construct the canonical boundary charges and the associated asymptotic symmetry algebra (ASA), including its central extension.

\subsection{Boundary conditions}\label{se:2.1}

We work with Eddington--Finkelstein type of coordinates, $x^{\mu} = (u,r,\vp)$, where $u$ is retarded time, $r$ the radial coordinate and $\vp\sim\vp+2\pi$ the angular coordinate. For convenience we employ a gauge where the radial dependence of the $\mathfrak{isl}(2)$ connection $A$ is captured by a group element $b \in ISL(2)$ which satisfies $\delta b =0$ and $b=b(r)$
\begin{equation}
A= b^{-1}(a + \extd\,)b\,.
\label{eq:bms4}
\end{equation}
Here the auxiliary connection $a$ is $r$-independent and has legs in the $(u, \vp)$-directions only
\begin{equation}
a = a_u(u,\vp) \extd u + a_{\vp}(u,\vp)\extd\vp\,.
\end{equation}

The loosest set of boundary conditions is then most easily parametrized by introducing six arbitrary state-dependent functions (also referred to as `vevs' or `normalizable modes' in a holographic context) in the $a_\vp$ component and six arbitrary chemical potentials (satisfying $\delta\mu = 0$; they are also referred to as `sources' or `non-normalizable modes' in a holographic context) in the $a_u$ component, i.e., one for each $\mathfrak{isl}(2)$ generator.
\begin{align}\label{abcs}
a_{\vp} & = - \left(\cM^+ L_+ - 2\cM^0 L_0 + \cM^- L_- + \cL^+ M_+ - 2 \cL^0 M_0 + \cL^- M_- \right) \\
a_u & = \mu_L^n L_n + \mu_M^n M_n
\label{eq:bms6}
\end{align}
This connection solves the flatness condition $\extd A + A \wedge A = 0$ if the state-dependent functions $\cM^a$ and $\cL^a$ satisfy
\begin{subequations}
\label{eq:bms5}
\begin{align}
\partial_u \cL^{\pm} & = \pm \mu_L^0 \cL^{\pm} \pm 2 \mu_L^{\pm} \cL^0 \pm \mu_M^0 \cM^{\pm} \pm 2 \mu_M^{\pm} \cM^0 - \partial_\vp \mu_M^{\pm} \\
\partial_u \cL^{0} & = \mu_L^+ \cL^{-} - \mu_L^{-} \cL^+ + \mu_M^+ \cM^{-} - \mu_M^{-} \cM^+ + \frac{1}{2}\partial_\vp \mu_M^{0} \\
\partial_u \cM^{\pm} & = \pm \mu_L^0 \cM^{\pm} \pm 2 \mu_L^{\pm} \cM^0 - \partial_\vp \mu_L^{\pm} \\
\partial_u \cM^{0} & = \mu_L^+ \cM^{-} - \mu_L^{-} \cM^+  + \frac12 \partial_\vp \mu_L^{0} \,.
\end{align}
\end{subequations}

Variations of the connection $A$ allowed by our boundary conditions above are given by
\eq{
\delta A = b^{-1} \delta a b = b^{-1}\, \sum_n \big(\cO(1) L_n + \cO(1) M_n\big) \, b\,\extd\vp\,.
}{eq:bms7}
In this section we do not specify the group element $b$, as it is not needed. We shall make a specific choice for $b$ in section \ref{se:4} when recovering our results in the metric formulation.

For later purposes we state here the length dimensions of relevant quantities appearing in our paper. Dimensionless:  $r$, $\vp$, $L_n$, $M_n$, ${\cal L}$, ${\cal M}$, $A$, $\extd A$, $e$, $\omega$, $a_\vp$, $b$, $I_{\textrm{\tiny CS}}$. Dimension 1: $u$, $G$, $\langle\rangle$. Dimension 2: $\widetilde{\tr}$. Dimension $-1$: $\mu_{L,\,M}$, $k$, $a_u$. This implies that some pure numbers in our paper actually have dimensions, but it is straightforward to determine them from these lists. For instance, the numbers in the matrix \eqref{eq:bi} have length dimension 1, and the numbers in the same matrix have length dimension 2 in the twisted trace \eqref{eq:ttr}.

\subsection{Boundary condition preserving transformations}\label{se:2.2}

We now look for all transformations that are locally pure gauge and that preserve the boundary conditions \eqref{eq:bms4}-\eqref{eq:bms7}, i.e. we look for gauge parameters 
\begin{equation}
\epsilon = b^{-1} (\epsilon_M^n M_n + \epsilon_L^n L_n) b\,,
\end{equation}
that satisfy 
\begin{equation}
\extd \epsilon + [A,\,\epsilon] = \cO(\delta A)
\end{equation}
with $\delta A$ as given in \eqref{eq:bms7}.

Since all state-dependence is in the $a_\vp$-component of the connection and it contains the most general form possible, there are no restrictions on the $\vp$-dependence of the gauge parameters. Thus the state-dependent functions transform as
\begin{subequations}\label{trafos}
\begin{align}
\delta \cL^{\pm} & = \pm \epsilon_L^0 \cL^{\pm} \pm 2 \epsilon_L^{\pm} \cL^0 \pm \epsilon_M^0 \cM^{\pm} \pm 2 \epsilon_M^{\pm} \cM^0 - \partial_\vp \epsilon_M^{\pm} \\
\delta \cL^{0} & = \epsilon_L^+ \cL^{-} - \epsilon_L^{-} \cL^+ + \epsilon_M^+ \cM^{-} - \epsilon_M^{-} \cM^+ + \frac{1}{2}\partial_\vp \epsilon_M^{0} \\
\delta \cM^{\pm} & = \pm \epsilon_L^0 \cM^{\pm} \pm 2 \epsilon_L^{\pm} \cM^0 - \partial_\vp \epsilon_L^{\pm} \\
\delta \cM^{0} & = \epsilon_L^+ \cM^{-} - \epsilon_L^{-} \cM^+  + \frac12 \partial_\vp \epsilon_L^{0} \,.
\end{align}
\end{subequations}
The $a_u$-component of the connection is state-independent according to our boundary conditions, i.e., $\delta a_u=0$.
This fixes the advanced time evolution of the gauge parameter $\epsilon$. 
\begin{subequations}
\begin{align}
 \partial_u \epsilon_M^{\pm} & =  \pm \epsilon_L^{\pm} \mu_M^0 \mp \epsilon_L^0 \mu_M^{\pm}  \pm \epsilon_M^{\pm} \mu_L^0 \mp \epsilon_M^0 \mu_L^{\pm} \\
\frac12 \partial_u \epsilon_M^{0} & = \epsilon_L^+ \mu_M^{-} - \epsilon_L^{-} \mu_M^+ + \epsilon_M^+ \mu_L^{-} - \epsilon_M^{-} \mu_L^+  \\
 \partial_u \epsilon_L^{\pm} & =  \pm \epsilon_L^{\pm} \mu_L^0 \mp \epsilon_L^0 \mu_L^{\pm} \\
\frac12 \partial_u \epsilon_L^{0} & = \epsilon_L^+ \mu_L^{-} - \epsilon_L^{-} \mu_L^+ 
\end{align}
\end{subequations}

\subsection{Canonical boundary charges and asymptotic symmetry algebra}\label{se:2.3}

The variation of the charge associated to an asymptotic gauge transformations is
\begin{equation}
\delta Q[\epsilon] = \frac{k}{2\pi} \oint \extd \vp \langle \epsilon \,\delta A_\vp \rangle
\label{eq:bms8}
\end{equation}
which in our case yields
\begin{equation}
\delta Q[\epsilon] = \frac{k}{\pi} \oint \extd \vp \left( \epsilon_M^+ \delta \cM^- + \epsilon_M^0 \delta\cM^0 + \epsilon_M^- \delta\cM^+ +\epsilon_L^+ \delta\cL^- + \epsilon_L^0 \delta\cL^0 + \epsilon_L^- \delta\cL^+ \right) \,.
\end{equation}
Assuming that the gauge parameter $\epsilon$ does not depend on any of the state-dependent functions $\cL^a, \cM^a$ one can trivially integrate the charges in field space.\footnote{
In more restrictive setups the boundary conditions may impose certain inter- and state-dependencies on the gauge parameters and integrability of the charges may become non-trivial. In the present case there is no restriction on the gauge parameters and hence it is possible to assume that they are all independent from each other and do not depend on the state. 
}
\begin{equation}
Q[\epsilon] = \frac{k}{\pi} \oint \extd \vp  \left( \epsilon_M^+ \cM^- + \epsilon_M^0 \cM^0 + \epsilon_M^- \cM^+ +\epsilon_L^+ \cL^- + \epsilon_L^0 \cL^0 + \epsilon_L^- \cL^+ \right) 
\label{eq:lalapetz}
\end{equation}

The ASA is spanned by the Poisson bracket algebra of the above charges, $\{Q[\epsilon_1], Q[\epsilon_2]\} = \delta_{\epsilon_1} Q[\epsilon_2]$. In terms of the Fourier modes
\begin{equation}\label{FM}
\cL_n^a = \frac{k}{\pi} \oint \extd\vp e^{-in\vp} \cL^a \qquad 
\cM_n^a = \frac{k}{\pi} \oint \extd\vp e^{-in\vp} \cM^a
\end{equation}
we obtain the algebra
\begin{subequations}
\begin{align}
\{\cL_n^a , \cL_m^b \} & = (a-b)\cL_{n+m}^{a+b} \\
\{\cL_n^a , \cM_m^b \} & = (a-b)\cM_{n+m}^{a+b} -  i n k \kappa_{ab}\delta_{n+m,0} \\
\{\cM_n^a , \cM_m^b \} & = 0\,.
\end{align}
\end{subequations}

Taking $M_n^a = i \cM_n^a, L_n^a = i \cL_n^a$ and replacing the Poisson brackets by commutators $i\{ \, , \} = [\, , ]$ yields the commutator algebra\footnote{%
The algebra \eqref{aff_isl2} can be obtained directly by a suitable \.In\"on\"u--Wigner contraction of two affine $\mathfrak{sl}(2)_k$ algebras. We consider such contractions arising from taking the large AdS radius limit in section \ref{se:ads}. Note that the central term in our algebra \eqref{aff_isl2} differs from the one obtained in earlier \.In\"on\"u--Wigner contractions of affine algebras \cite{Majumdar:1992fp}.
}
\begin{subequations}
\label{aff_isl2}
\begin{align}
[L_n^a , L_m^b] & = (a-b)L_{n+m}^{a+b} \\
[L_n^a , M_m^b] & = (a-b)M_{n+m}^{a+b} -  n k \kappa_{ab}\delta_{n+m,0} \\
[M_n^a , M_m^b] & = 0\,.
\end{align}
\end{subequations}
This is the affine $\mathfrak{isl}(2)_k$ algebra. Like in the AdS$_3$ case it may have been anticipated that the loosest set of boundary conditions leads to the loop algebra of the underlying gauge algebra as ASA \cite{Elitzur:1989nr}. Our result \eqref{aff_isl2} confirms this expectation.

\section{Metric formulation}\label{se:4}

\subsection{Boundary conditions in generalized Fefferman--Graham gauge}

The gauge choice \eqref{eq:bms4} is convenient for the analysis in Chern--Simons theory, but in order to connect to the metric formulation we have to specify a suitable group element $b(r)$. Suitable means in this respect that the same number of independent functions appearing in the gauge connection \eqref{eq:bms6} should appear in the metric \cite{Grumiller:2016pqb}. A specific choice for $b(r)$ that satisfies this property is
    \begin{equation}
        b=e^{r L_0}e^{\frac{1}{2}(M_{+1}+M_{-1})}\,. 
    \end{equation}    
    
Using this $b$ yields twelve free functions in the dreibein and in the metric. The resulting metric has the form of a generalized Fefferman--Graham expansion \cite{Grumiller:2016pqb},\footnote{%
Since expressions like $e^r$ may look confusing on dimensional grounds we recall that $r$ is dimensionless, but nevertheless $\extd s^2$ has length dimension 2, see the list of length dimensions below equation \eqref{eq:bms7}.
}
    \begin{align}\label{genFG}
        \extd s^2 =  \extd r^2+\Big(e^{r}\, N_i^{(0)} + \, N_i^{(1)} & + e^{-r}N_i^{(2)}  \Big)\extd r \, \extd x^i \nonumber \\
        + & \Big(e^{2r}\,g_{ij}^{(0)} + e^{r}\, g_{ij}^{(1)} + g_{ij}^{(2)} + {\cal O}(e^{-r})\Big)\, \extd x^i\extd x^j\,,
    \end{align}
where all expansion coefficients $g_{ij}^{(n)}$ depend on the boundary coordinates $x^i=(u,\vp)$, only. Our boundary conditions on the metric are then summarized in the next three sets of equations. The shift vector components $N_i^{(n)}$ are given by
\begin{subequations}
 \label{eq:bcN}
 \begin{align}
  N_u^{(0)} = g_{ur}^{(0)} &= - \mu_M^+ &
  N_\vp^{(0)} = g_{\vp r}^{(0)} &= \cL^+ \\
  N_u^{(1)} = g_{ur}^{(1)} &= \mu_L^0 & 
  N_\vp^{(1)} = g_{\vp r}^{(1)} &= 2\cM^0 \\
  N_u^{(2)} = g_{ur}^{(2)} &= \mu_M^- &
  N_\vp^{(2)} = g_{\vp r}^{(2)} &= -\cL^- \,.
 \end{align}
\end{subequations}
The diagonal metric components $g_{ij}^{(n)}$ have a similar structure as the shift-vector components regarding the sources $\mu_L, \mu_M$ and vevs $\cL, \cM$.
\begin{subequations}
 \label{eq:bcg}
 \begin{align}
  g_{uu}^{(0)} &= (\mu_L^+)^2 & 
  g_{\vp\vp}^{(0)} &= (\cM^+)^2 \\
  g_{uu}^{(1)} &= 2\mu_L^+ \mu_M^0 - 2\mu_L^0\mu_M^+ & 
  g_{\vp\vp}^{(1)} &= 4\cL^+ \cM^0 - 4\cM^+\cL^0 \\
  g_{uu}^{(2)} &= (\mu_L^0)^2\!+(\mu_M^0)^2\!-2\mu_L^+\mu_L^-\!-4\mu_M^+\mu_M^- & 
  g_{\vp\vp}^{(2)} &= 4(\cL^0)^2\!+4(\cM^0)^2\!-4\cL^+\cL^-\!-2\cM^+\cM^-
 \end{align}
\end{subequations}
The off-diagonal metric components $g_{u\vp}$ are not independent from the expressions above, but determined by them algebraically through the Einstein equations.
\begin{subequations}
 \label{eq:gtphi}
 \begin{align}
  g_{u\vp}^{(0)} &=  - \cM^+\mu_L^+ \\
  g_{u\vp}^{(1)} &=  \cL^+\mu_L^0 - \cM^+\mu_M^0 + 2\cL^0\mu_L^+ - 2\cM^0\mu_M^+ \\
  g_{u\vp}^{(2)} &=  2\cM^0\mu_L^0 + 2\cL^0\mu_M^0 + 2\cL^-\mu_M^+ + 2\cL^+\mu_M^- + \cM^+\mu_L^- + \cM^-\mu_L^+ 
 \end{align}
\end{subequations}
In addition to these relations, demanding the metric to be Ricci-flat imposes the on-shell constraints \eqref{eq:bms5}. 
Up to these on-shell constraints, which are different for AdS$_3$, we have recovered exactly the same asymptotic form \eqref{genFG}-\eqref{eq:gtphi} as in the AdS$_3$ construction \cite{Grumiller:2016pqb}.
%

\subsection{Asymptotic Killing vectors}\label{se:3.3}

The asymptotic Killing vectors preserving our boundary conditions \eqref{genFG}-\eqref{eq:gtphi} [subject to \eqref{eq:bms5}] are all vector fields $\xi^\mu$ that satisfy the asymptotic Killing equation
\eq{
{\cal L}_\xi g_{\mu\nu} = {\cal O}(\delta g_{\mu\nu})
}{eq:Lie}
where ${\cal L}_\xi$ denotes the Lie-variation along $\xi$ and $\delta g_{\mu\nu}$ are the variations allowed by our boundary conditions \eqref{genFG}-\eqref{eq:gtphi}. Writing
\eq{
\xi^\mu(u,\,\vp,\,r) = \xi^\mu_{(0)}(u,\,\vp) + e^{-r}\,\xi^\mu_{(1)}(u,\,\vp) + e^{-2r}\,\xi^\mu_{(2)}(u,\,\vp) + {\cal O}(e^{-3r})
}{eq:xiAnsatz}
we can obtain the asymptotic Killing vectors satisfying \eqref{eq:Lie} from solving the relation $\epsilon = e_{\mu}\xi^{\mu}$. 
This leads to the result
\begin{subequations}
 \label{eq:xi}
 \begin{align}
  \xi^\vp_{(0)} &=- \frac{\mu^+_L\epsilon_M^+-\mu^+_M\epsilon_L^+}{\mathcal{L}^+\mu^+_L-\mathcal{M}^+\mu_M^+} &
  \xi^u_{(0)} &= \frac{\mathcal{L}^+\epsilon_L^+-\mathcal{M}^+\epsilon_M^+}{\mathcal{L}^+\mu^+_L-\mathcal{M}^+\mu_M^+} \\
  \xi^\vp_{(1)} &= \frac{\mu^+_M\lambda^0}{\mathcal{L}^+\mu^+_L-\mathcal{M}^+\mu_M^+} &
  \xi^u_{(1)} &= \frac{\mathcal{L}^+\lambda^0}{\mathcal{L}^+\mu^+_L-\mathcal{M}^+\mu_M^+} \\
  \xi^\vp_{(2)} &= -\frac{\mu^+_L\lambda^+_M-\mu^+_M\lambda^+_L}{\mathcal{L}^+\mu^+_L-\mathcal{M}^+\mu_M^+}&
  \xi^u_{(2)} &= \frac{\mathcal{L}^+\lambda^+_L-\mathcal{M}^+\lambda^+_M}{\mathcal{L}^+\mu^+_L-\mathcal{M}^+\mu_M^+}
 \end{align}
 \begin{align}
   \xi^\rho_{(0)} &= \epsilon_L^0-2\mathcal{M}^0\xi^\vp_{(0)}-\mu_L^0\xi^u_{(0)} \\
   \xi^\rho_{(1)} &= 2\mathcal{L}^-\xi^\vp_{(0)}-2\mathcal{M}^0\xi^\vp_{(1)}+2\epsilon_M^- -\mu_L^0\xi^u_{(1)}-2\mu_M^-\xi^u_{(0)}\\
   \xi^\rho_{(2)} &= 2\mathcal{L}^-\xi^\vp_{(1)}-2\mathcal{M}^0\xi^\vp_{(2)}-\mu_L^0\xi^u_{(2)}-2\mu_M^-\xi^u_{(1)}
 \end{align}
with
\begin{align}
		\lambda^0 & =\epsilon_M^0-\mu_M^0\xi^u_{(0)}-2\mathcal{L}^0\xi^\vp_{(0)}\qquad\qquad \lambda_M^+=\tfrac{1}{2}\left(\xi^r_{(1)}+\mu_L^0\xi^u_{(1)}+2\mathcal{M}^0\xi^\vp_{(1)}\right)\\
		\lambda_L^+ & = \mu_L^-\xi^u_{(0)}-\mu_M^0\xi^u_{(1)}-\mathcal{M}^+\xi^\vp_{(0)}-\epsilon_L^--2\mathcal{L}^0\xi^\vp_{(1)}.
	\end{align}
\end{subequations}
Here $\epsilon_L^a(u,\vp)$ and $\epsilon_M^a(u,\vp)$ with $a=0,\pm$ denote six arbitrary free functions. The asymptotic Killing vectors are state-dependent even to leading order. As in the AdS case \cite{Grumiller:2016pqb} this state-dependence is crucial for obtaining the correct ASA.

The usual procedure when determining the ASA involves evaluating the Lie bracket between the asymptotic Killing vectors
	\begin{equation}\label{eq:UsualLieBracket}
		[\xi_1,\xi_2]^\mu=\mathcal{L}_{\xi_1}\xi_2^\mu\,.
	\end{equation}
However, the expression \eqref{eq:UsualLieBracket} is only valid if the relevant pieces of the asymptotic Killing vectors do not depend on state-dependent functions. Thus one has to modify \cite{Barnich:2010eb} (or ``adjust'' \cite{Compere:2015knw}) the Lie bracket \eqref{eq:UsualLieBracket} as follows   
	\begin{equation}\label{eq:ModLieBracket}
		[\xi_1,\xi_2]^\mu_M=\mathcal{L}_{\xi_1}\xi_2^\mu-\delta^g_{\xi_1}\xi_2^\mu+\delta^g_{\xi_2}\xi_1^\mu\,,
	\end{equation}
where $\delta^g_{\xi_1}\xi_2^\mu$ denotes the change induced in $\xi_2^\mu(g)$ due to the variation $\delta^g_{\xi_1}g_{\mu\nu}=\mathcal{L}_{\xi_1}g_{\mu\nu}$.
A straightforward but tedious calculation using the modified Lie bracket \eqref{eq:ModLieBracket} yields 
	\begin{equation}
		[\xi(\{\epsilon_L^{a;1},\epsilon_M^{a;1}\}),\xi(\{\epsilon_L^{a;2},\epsilon_M^{a;2}\})]^\mu=\xi^\mu(\{\epsilon_L^{a;[1,2]},\epsilon_M^{a;[1,2]}\})\,,
	\end{equation}
with
	\begin{subequations}
	\begin{align}
		\epsilon^{\pm;[1,2]}_L&=\pm\epsilon_L^{0;1}\epsilon^{\pm;2}_L\mp\epsilon^{0;2}_L\epsilon^{\pm;1}_L\qquad
		\epsilon^{0;[1,2]}_L=2\left(\epsilon^{-;1}_L\epsilon^{+;2}_L-\epsilon^{-;2}_L\epsilon^{+;1}_L\right)\\
		\epsilon_M^{\pm;[1,2]}&=\pm\epsilon_L^{0;1}\epsilon^{\pm;2}_M\mp\epsilon^{0;2}_L\epsilon^{\pm;1}_M\mp\epsilon_L^{\pm;1}\epsilon^{0;2}_M\pm\epsilon^{\pm;2}_L\epsilon^{0;1}_M\\
		\epsilon^{0;[1,2]}_M &=2\left(\epsilon^{-;1}_L\epsilon^{+;2}_M-\epsilon^{-;2}_L\epsilon^{+;1}_M-\epsilon^{+;1}_L\epsilon^{-;2}_M+\epsilon^{+;2}_L\epsilon^{-;1}_M\right)
	\end{align}
	\end{subequations}
and $\{\epsilon_L^a,\epsilon_M^a\}=(\epsilon_L^+,\epsilon_L^0,\epsilon_L^-,\epsilon_M^+,\epsilon_M^0,\epsilon_M^-)$.
Introducing Fourier modes ($e_{n|m}:=e^{in\varphi+imu}$)
	\begin{subequations}
	\begin{align}
		L^+_{n|m}&=\xi^\mu(0_2,e_{n|m},0_3) & L^0_{n|m}&=\xi^\mu(0,e_{n|m},0_4) & L^-_{n|m}&=\xi^\mu(e_{n|m},0_5)\\
		M^+_{n|m}&=\xi^\mu(0_5,e_{n|m}) & M^0_{n|m}&=\xi^\mu(0_4,e_{n|m},0) & M^-_{n|m}&=\xi^\mu(0_3,e_{n|m},0_2)
	\end{align}
	\end{subequations}
where $0_n$ denotes $n$ zeros (e.g.~$0_3=0,0,0$), one finds that these modes satisfy
	\begin{subequations}\label{eq:AKVAlgebraFromCS}
	\begin{align}
		[L^a_{n|p},L^b_{m|q}]&=(a-b)L^{a+b}_{n+m|p+q}\\
		[L^a_{n|p},M^b_{m|q}]&=(a-b)M^{a+b}_{n+m|p+q}\\
		[M^a_{n|p},M^b_{m|q}]&=0
	\end{align}
	\end{subequations}
which is essentially \eqref{aff_isl2} but without the central extensions, and with a double Fourier expansion with respect to $u$ and $\vp$. 
This shows the consistency of the algebra of asymptotic Killing vectors with the canonical realization of the ASA \eqref{aff_isl2}.

\section{Special cases}\label{se:5}

In this section we consider special cases by suitably restricting some of the state-dependent functions and chemical potentials.

In section \ref{se:5.1} we recover Barnich--Comp\`ere boundary conditions and $\mathfrak{bms}_3$.
In section \ref{se:5.2} we get Heisenberg boundary conditions.
In section \ref{se:5.3} we obtain the Detournay--Riegler generalization of $\mathfrak{bms}_3$.

\subsection{$\mathfrak{bms}_3$}\label{se:5.1}

The general boundary conditions of the last section reduce to the known flat space boundary conditions of \cite{Barnich:2006av} when restricting the state-dependent functions as
\begin{equation}\label{BMSconstraints}
\cL^0 = 0 = \cL^+ = \cM^0\,, \qquad \cM^+ = -1\,, \qquad \cM^-, \cL^- = \text{arbitrary}\,.
\end{equation}
The field equations $F=0$ then fix four of the state-independent chemical potentials.
\begin{subequations}
\begin{align}
\mu_M^0 & = - \partial_{\vp}\mu_M^+  & 
\mu_M^- & = - \frac12 \partial_{\vp}^2  \mu_M^+ - \cL^- \mu_L^+ - \cM^- \mu_M^+
\\
\mu_L^0 & = - \partial_{\vp}\mu_L^+ &
\mu_L^- & = - \frac12 \partial_{\vp}^2  \mu_M^+ - \cM^- \mu_L^+
\end{align}
\end{subequations}
The retarded time evolution of the two arbitrary state-dependent functions is given by
\begin{subequations}
\label{eq:angelinajolie}
\begin{align}
\partial_u \cM^- & = \frac12\partial_{\vp}^3 \mu_M^+ + 2\partial_{\vp} \mu_L^+ \cM^- + \mu_L^+ \partial_{\vp}\cM^- \\ 
\partial_u \cL^- & = \frac12\partial_{\vp}^3 \mu_M^+ + 2\partial_{\vp} \mu_L^+ \cL^- + \mu_L^+ \partial_{\vp}\cL^- + 2\partial_{\vp} \mu_M^+ \cM^- + \mu_M^+ \partial_{\vp}\cM^-\,.
\end{align}
\end{subequations}
From the right hand side of \eqref{eq:angelinajolie} we recognize the transformation properties of the $\mathfrak{bms}_3$ algebra. Indeed we find that the asymptotic symmetry analysis in this case gives the $\mathfrak{bms}_3$ algebra. Equivalently, this can also be seen by reducing the Poisson brackets of the affine $\mathfrak{isl}(2)_k$ algebra to the Dirac bracket on the constraint surface defined by \eqref{BMSconstraints}, as we will now show.

In terms of the Fourier modes of the charges \eqref{FM} the constraints \eqref{BMSconstraints} read
\begin{equation}
L_n^+ = L_n^0 =M_n^0 = 0\,, \qquad M_n^+ = - k \delta_{n,0}\,,
\end{equation}
and the matrix of Poisson brackets in the basis $\{L_n^+, L_n^0, M_n^+, M_n^0 \}$ on this constraint surface is given by
\begin{equation}
C_{\alpha\beta} = \left(\begin{array}{cccc} 
0	 	& 		0 		& -k \delta_{n+m}	& 	0 		\\
0		&		0		& -\frac{k}{2}n\delta_{n+m} &	k\delta_{n+m} 	\\
k\delta_{n+m}	& -\frac{k}{2}n\delta_{n+m}	& 	0 		&	0		\\
0		& - k \delta_{n+m}		&	0		&	0
\end{array}\right)\,.
\end{equation}
This matrix is invertible and its inverse can be used to compute the Dirac brackets
\begin{equation}
[a,b]^* = \{a,b\} - \{a, \chi_\alpha\}C^{\alpha\beta}\{\chi_{\beta}, b\}\,.
\end{equation}
Explicit computation shows that on the constraint surface defined by \eqref{BMSconstraints} the Dirac brackets reduce to the $\mathfrak{bms}_3$ algebra as expected. Defining $M_n := M_n^-$ and $L_n := - L_n^-$ we find
\begin{subequations}
\label{bms3}
\begin{align}
[L_n, L_m]^* & = (n-m)\cL_{n+m}  \\
[L_n, M_m]^* & = (n-m)\cM_{n+m} + k n^3\delta_{n+m}\\
[M_n, M_m]^* & = 0\,. 
\end{align}
\end{subequations}
The algebra \eqref{bms3} coincides precisely with the centrally extended $\mathfrak{bms}_3$ algebra discovered in \cite{Barnich:2006av}.

\subsection{Heisenberg}\label{se:5.2}

The second known special case we discuss is the reduction to two affine $\hat{\mathfrak{u}}(1)$ algebras or, equivalently, the Heisenberg algebra with two central Casimirs first described in \cite{Afshar:2016wfy} in the context of AdS$_3$ Einstein gravity. The constraints of the state-dependent functions in this case read
\begin{equation}\label{remembermyname}
\cM^{\pm} = 0 = \cL^{\pm} \qquad \cM^0, \cL^0 = \text{arbitrary}\,.
\end{equation}
The field equations now reduce to
\begin{equation}
\partial_u\cL^0 = - \frac12 \partial_{\vp} \mu_M^0 \qquad
\partial_u\cM^0 = - \frac12 \partial_{\vp} \mu_L^0\,.
\end{equation}
We fixed partly the gauge such that $\mu_M^{\pm}$ and $\mu_L^{\pm}$ all vanish, as these chemical potentials no longer play a role in the asymptotic charges nor their transformation properties.

Reducing the Poisson bracket algebra to the constraint surface defined by \eqref{remembermyname} essentially gives the subalgebra spanned by the Fourier modes of $\cL^0$ and $\cM^0$
\begin{equation}\label{heisenberg}
[L_n^0, M_n^0] = - k n \delta_{n+m}\, \qquad [L_n^0,L_m^0] = 0 = [L_n^0,L_m^0]\,.
\end{equation}
Or, after taking $J^{\pm}_{\pm n} = \frac{1}{2}\left( L_n^0 \pm M_n^0 \right)$, the algebra becomes that of two affine $\hat{\mathfrak{u}}(1)$ currents as in \cite{Afshar:2016wfy, Afshar:2016kjj}.
\begin{equation}
[J_n^\pm, J_n^\pm] = \pm \frac{k}{2} n \delta_{n+m}
\end{equation}

\subsection{Detournay--Riegler}\label{se:5.3}

The third special case we discuss is the Drinfeld-Sokolov reduction to a $\mathfrak{bms}_3$ algebra and two affine $\hat{\mathfrak{u}}(1)$ current algebras first described in \cite{Detournay:2016sfv}. These boundary conditions can be obtained from our general boundary restrictions by setting
    \begin{equation}
        \mathcal{M}^-=\frac{\mathcal{M}}{4}e^{-\alpha},\quad\mathcal{M}^+=-e^{\alpha},\quad \mathcal{L}^-=\frac{e^{-\alpha}}{4}\left(2\mathcal{N}-\beta\mathcal{M}\right),\quad\mathcal{L}^+=-\beta e^{\alpha},
    \end{equation}
and
    \begin{equation}
        \mathcal{M}^0=\mathcal{L}^0=0,\qquad \mathcal{N},\,\mathcal{M},\,\alpha,\,\beta=\textrm{arbitrary}.
    \end{equation}
In addition the chemical potentials have to be fixed to
    \begin{equation}
        \mu_M^+=e^{\alpha},\qquad\mu_M^-=-\frac{\mathcal{M}}{4}e^{-\alpha},\qquad \mu_M^0=\mu_L^a=0,
    \end{equation}
where by the equations of motion $F=0$ one has
    \begin{subequations}\label{eq:StateOnShell}
    \begin{align}
    \mathcal{M} &=  \mathcal{M}(\varphi)& \mathcal{N} &= \mathcal{L}(\varphi) +\frac{u}{2}\mathcal{M}' \\
    \alpha &= A(\varphi)&\beta &= B(\varphi) + u A'
    \end{align}
    \end{subequations}
and primes denote derivatives with respect to $\varphi$. 

For this choice of boundary conditions the boundary condition preserving transformations reduce to
	\begin{subequations}\label{eq:BCPGTs}
	\begin{align}
		\epsilon^1_L&=e^\alpha\epsilon_\mathcal{L}\,,\quad \epsilon^0_L=\epsilon_\mathcal{J}\,,\quad
		\epsilon^{-1}_L=-\frac{e^{-\alpha}}{4}\left(2\epsilon_\mathcal{J}'+\mathcal{M}\epsilon_\mathcal{L}\right)\,,\\
		\epsilon^1_M&=e^\alpha\left(\beta\epsilon_\mathcal{L}+\sigma_\mathcal{M}\right),\quad\epsilon^0_M=\sigma_\mathcal{P}\,,\\
		\epsilon^{-1}_M&=\frac{e^{-\alpha}}{2}\left[\frac{\mathcal{M}}{2}\left(\beta\epsilon_\mathcal{L}-\sigma_\mathcal{M}\right)-\mathcal{N}\epsilon_\mathcal{L}+\beta\epsilon_\mathcal{J}'-\sigma_\mathcal{P}'\right]\,,
	\end{align}
where
    \begin{equation}
    \epsilon_\mathcal{L} =  \epsilon_\mathcal{L}(\varphi)\qquad\epsilon_\mathcal{J} = \epsilon_\mathcal{J}(\varphi)\qquad \sigma_\mathcal{M} = \epsilon_\mathcal{M}(\varphi) +u\,\epsilon_\mathcal{L}'\qquad\sigma_\mathcal{P} = \epsilon_\mathcal{P}(\varphi) + u\,\epsilon_\mathcal{J}'\,.
    \end{equation}
	\end{subequations}
The redefinitions
	\begin{equation}
		\tilde{\mathcal{M}}=\frac{k}{4\pi}\mathcal{M}\qquad\tilde{\mathcal{L}}=\frac{k}{2\pi}\mathcal{L}\qquad\mathcal{P}=-\frac{k}{2\pi}A'\qquad\mathcal{J}=-\frac{k}{2\pi}B'
	\end{equation}
simplify the canonical boundary charges to
    \begin{equation}
		Q[\epsilon] =\oint \extd\varphi\left(\tilde{\mathcal{M}}\epsilon_\mathcal{M}+\tilde{\mathcal{L}}\epsilon_\mathcal{L}+\mathcal{P}\epsilon_\mathcal{P}+\mathcal{J}\epsilon_\mathcal{J}\right).
	\end{equation}
Introducing suitable Fourier modes the corresponding ASA is then essentially\footnote{In order to obtain this algebra a (twisted) Sugawara shift of $\tilde{\mathcal{L}}$ and $\tilde{\mathcal{M}}$ has to be implemented along the same lines as in \cite{Detournay:2016sfv}.} 
given by
	\begin{subequations}\label{eq:ASAMiddle}
	\begin{align}
		[L_n,L_m]&=(n-m)L_{n+m}\label{eq:VirasoroCentral}\\
		[L_n,M_m]&=(n-m)M_{n+m}+\frac{c_M}{12}n^3\delta_{n+m,0}\\
		[L_n,J_m]&=-mJ_{n+m}\\
		[L_n,P_m]&=-mP_{n+m}\\
		[M_n,J_m]&=-mP_{n+m}\\
		[J_n,P_m]&=\kappa_P n\,\delta_{n+m,0}\label{eq:AffU1}
	\end{align}
	\end{subequations}
where $c_M=12k$ and $\kappa_P=-k$. 
Note that in this special case the chemical potentials (and subsequently also the gauge parameters) depend on the state-dependent functions. Hence we have to relax the assumption that $\delta a_u = 0$ made in section \ref{se:3}. However, the resulting boundary conditions are consistent and can be regarded as a special case of our more general boundary conditions, but with state-dependent chemical potentials.

\section{Small cosmological constant limits of AdS}\label{se:ads}

The general flat space boundary conditions and the special cases of the preceding section can be obtained from the most general AdS$_3$ boundary conditions of \cite{Grumiller:2016pqb} by means of a flat space limit, where the cosmological constant is sent to zero (or equivallently, the AdS radius is sent to infinity). In this section we discuss this limit. Besides recovering known results we find new flat limits leading to Carroll gravity in three dimensions.

In section \ref{se:bh} we recapitulate the results for possible ASAs coming from a Brown--Henneaux type of analysis for AdS$_3$ Einstein gravity.
In section \ref{se:csc} we study chirally symmetric contractions of AdS$_3$ symmetry algebras to their corresponding flat space counterparts.
In section \ref{se:5.5} we discover two novel types of boundary conditions for Carroll gravity by virtue of a chirally symmetric and a non-chiral contraction from AdS$_3$ Einstein gravity.

\subsection{Recap of Brown--Henneauxlogy of AdS$_3$}\label{se:bh}

The generic AdS$_3$ boundary conditions of \cite{Grumiller:2016pqb} contain six arbitrary state-dependent functions and six chemical potentials, three of both in each chiral sector. The ASA for the most generic set of boundary conditions is the loop group of $\mathfrak{sl}(2,\mathbb{R})_L \oplus \mathfrak{sl}(2,\mathbb{R})_R$, viz., the affine $\mathfrak{sl}(2)_k \oplus \mathfrak{sl}(2)_k$ algebra. Special cases of these boundary conditions are obtained by imposing conditions on the state-dependent functions and chemical potentials, although not all of these constraints will lead to consistent boundary conditions, as it may happen that the charges are not integrable. Consistent sets of constraints were discussed as special cases in \cite{Grumiller:2016pqb} and they are summarized in table \ref{table1} for one chiral sector, with the $\mathfrak{sl}(2,\mathbb{R})$ Chern--Simons connection $A$ parameterized as
\begin{align}
A & = b^{-1}(a + \extd\,)b \\
a_{\vp} & = - \mathfrak{L}^+ L_+ + 2 \mathfrak{L}^0 L_0 + \mathfrak{L}^- L_- \\
a_t & = \mu^+ L_+ + \mu^0 L_0 + \mu^- L_-\,.
\end{align}

\begin{table}[]
    \centering
    \begin{tabular}{c|c|c|c}
    & free functions & constraints    & ASA \\ \hline
    case 1 & 3   &   none    & $\mathfrak{sl}(2)_k$ \\
    case 2 & 2   & $\mathfrak{L}^0 = 0\,, \mathfrak{L}^+ = - e^{2\phi(x^+)}\,, \mathfrak{L}^- = e^{-2\phi(x^+)}\cL(x^+)$ & Vir$\oplus \mathfrak{u}(1)_k$ \\
    case 3 & 1   & $\mathfrak{L}^0 = 0\,, \mathfrak{L}^+ = -1$ & Vir \\
    case 4 & 1   & $\mathfrak{L}^{\pm} = 0$ & $\mathfrak{u}(1)_k$ \\
    case 5 & 0   &  $\mathfrak{L}^0 = \mathfrak{L}^\pm = 0$ & trivial
    \end{tabular}
    \caption{Symmetry algebra menagerie for AdS$_3$ Einstein gravity}
    \label{table1}
\end{table}
The combination of these boundary conditions with the other chiral sector of the theory leads to fourteen\footnote{There are fifteen inequivalent ways to combine cases 1-5 in left and right sectors, but we exclude the degenerate (case 5, case 5) scenario.} possible consistent boundary conditions from this (possibly non-exhaustive) list in table \ref{table1}. These contain the known cases of the Brown--Henneaux boundary conditions \cite{Brown:1986nw} (case 3, case 3), the Troessaert boundary conditions \cite{Troessaert:2013fma} (case 2, case 2), the Heisenberg boundary conditions \cite{Afshar:2016wfy} (case 4, case 4), the Comp{\`e}re--Song--Strominger boundary conditions \cite{Compere:2013bya} (case 3, case 4) or the Avery--Poojary--Suryanarayana boundary conditions \cite{Avery:2013dja} (case 1, case 3). However, in principle any combination of these constraints defines a consistent set of boundary conditions for AdS$_3$ gravity, not all of which have been discussed in the literature so far. If one allows for dependencies of chemical potentials on the charges further possibilities arise not captured by table \ref{table1}, see our discussion in section \ref{se:6.4}.

In the following we discuss the flat space limit of chirally symmetric combinations of these special cases, which lead to the special cases described in the last section. Afterwards we discuss a slightly different type of large AdS-radius limit for a chirally symmetric and a chirally non-symmetric combination and find two novel sets of boundary conditions for Carroll gravity in this way.

\subsection{Chirally symmetric contractions}\label{se:csc}

The flat space limit of AdS$_3$ gravity is most easily implemented at the level of the Chern--Simons gauge group. The AdS$_3$ isometry algebra $\mathfrak{sl}(2,\mathbb{R})\oplus \mathfrak{sl}(2,\mathbb{R})$ with generators $L^{\pm}_n$ for $n = -1,0,+1$ gives the three dimensional Poincar{\'e} algebra when contracted as follows in the large $\ell$ limit.
\begin{equation}
    M_n = \frac{1}{\ell} (L^+_n + L^-_{-n})  \qquad 
    L_n = (L^+_n - L_{-n}^-)
\end{equation}
For the most general boundary conditions, with no constraints on the state-dependent functions, this implies a limit from the generators $\mathfrak{J}^a_n$ and $\bar{\mathfrak{J}}^a_n$ of  $\mathfrak{sl}(2)_{\hat{k}}\oplus\bar{\mathfrak{sl}}(2)_{\hat{k}}$ to the  $\mathfrak{isl}(2)_k$ generators $L^a_n$ and $M_n^a$
\begin{equation}\label{limit_gen}
    M_n^a = \frac{1}{\ell}\left(\mathfrak{J}_n^a + \bar{\mathfrak{J}}_{-n}^{-a} \right)  \qquad
    L_n^a = \mathfrak{J}_n^a - \bar{\mathfrak{J}}_{-n}^{-a}\,.
\end{equation}
This redefinition yields the algebra \eqref{aff_isl2} with $k = \hat{k}/\ell$, where $\hat{k}$ is the level of the affine $\mathfrak{sl}(2)_{\hat{k}}$ algebras, one of which we display here,
\eq{
[\mathfrak{J}_n^a,\,\mathfrak{J}_m^b] = (a-b)\mathfrak{J}_{n+m}^{a+b} - \frac{1}{2} n\hat k\,\kappa_{ab}\,\delta_{n+m,\,0}\,.
}{eq:sl2}

The flat space limit \eqref{limit_gen} combines the left and right moving sectors symmetrically (while exchanging the $a=\pm1$ with $a=\mp1$ generators in the right moving sector). This implies that flat limits of all the chirally symmetric contractions are well-defined, under the provision that in the right moving sector the state-dependent functions $\bar{\mathfrak{L}}^n$ with $n=+1$  and $n=-1$ are exchanged with respect to the left moving sector.

\subsubsection{$\mathfrak{bms}_3$ limit}
The $\mathfrak{bms}_3$ algebra \eqref{bms3} is obtained from the Brown--Henneaux boundary conditions $\mathfrak{L}^0 = 0\,, \mathfrak{L}^+ = -1$ and $\bar{\mathfrak{L}}^0 = 0\,, \bar{\mathfrak{L}}^- = -1$. This leads to two copies of the Virasoro algebra with generators $\mathfrak{J}^-_n$ and $\bar{\mathfrak{J}}^+_n$ as the Fourier modes of $\mathfrak{L}^-$ and $\bar{\mathfrak{L}}^+$. The limit from two Virasoro generators to the $\mathfrak{bms}_3$ algebra is in these conventions given by \cite{Barnich:2012aw}
\begin{equation}\label{limit_bms}
    M_n = \frac{1}{\ell}\left(\mathfrak{J}_n^{-} + \bar{\mathfrak{J}}_{-n}^{+} \right)  \qquad
    L_n = \mathfrak{J}_n^{-} - \bar{\mathfrak{J}}_{-n}^{+}\,.
\end{equation}

\subsubsection{$\mathfrak{u}(1) \oplus \mathfrak{u}(1)$ limit} 
The near-horizon boundary conditions of \cite{Afshar:2016wfy} in AdS$_3$ gravity lead to a $\mathfrak{u}(1)_{\hat{k}} \oplus \mathfrak{u}(1)_{\hat{k}}$ algebra. These are implemented by considering case 4 in both chiral sectors. The constraints are $\mathfrak{L}^{\pm} = 0 = \bar{\mathfrak{L}}^{\mp}$ and the limit to \eqref{heisenberg} follows from
\begin{equation}\label{limit_u1}
    M^0_n = \frac{1}{\ell}\left(\mathfrak{J}_n^{0} + \bar{\mathfrak{J}}_{-n}^{0} \right)  \qquad
    L^0_n = \mathfrak{J}_n^{0} - \bar{\mathfrak{J}}_{-n}^{0}\,.
\end{equation}

\subsubsection{Extended $\mathfrak{bms}_3$ limit} 
The boundary conditions of \cite{Detournay:2016sfv} described in section \ref{se:5.3} can also be obtained as a limit from the boundary conditions of Troesseart presented in \cite{Troessaert:2013fma}. These boundary conditions lead to an ASA which is essentially two copies of a Vir$\oplus\mathfrak{u}(1)_k$ algebra. For each chiral sector we have
\begin{subequations}
\begin{align}
    [\mathfrak{L}_n,\mathfrak{L}_m] & = (n-m)\mathfrak{L}_{n+m} + \frac{c}{12}n^3 \delta_{m+n,0}\,, \\
    [\mathfrak{L}_n,\mathfrak{J}_m] & = -m\mathfrak{J}_{n+m}\,, \\
    [\mathfrak{J}_n,\mathfrak{J}_m] & = - \frac{\hat{k}}{2} n \delta_{n+m,0}\,. 
\end{align}
\end{subequations}
Here $\hat{k} = \ell/4G$ and $c=6\hat{k}$ and the generators of the other chiral sector are as usual denoted by barred quantities. The symmetry algebra \eqref{eq:ASAMiddle} is obtained by taking
\begin{equation}\label{limit_ebms}
    M_n = \frac{1}{\ell}\left(\mathfrak{L}_n + \bar{\mathfrak{L}}_{-n} \right) \quad\;
    L_n = \mathfrak{L}_n - \bar{\mathfrak{L}}_{-n} \quad\;
    P_n = \frac{1}{\ell}\left(\mathfrak{J}_n + \bar{\mathfrak{J}}_{-n} \right) \quad\;
    J_n = \mathfrak{J}_n - \bar{\mathfrak{J}}_{-n} \,.
\end{equation}

\subsection{New Carroll gravity boundary conditions from flat limit of AdS}\label{se:5.5}

An interesting question is whether it is possible to combine a flat limit from some AdS$_3$ boundary conditions with an additional contraction, like an ultra-relativistic limit to Carroll gravity. In this section we discuss two such examples, one that starts from a chirally symmetric AdS$_3$ configuration and one that starts from a chirally non-symmetric one. In this way we establish two novel sets of boundary conditions for Carroll gravity.

\subsubsection{Carroll gravity loop group boundary conditions}

\newcommand{\mJ}{\mathfrak{J}}

Starting in AdS$_3$ with case 1 in both sectors (see table \ref{table1}) as a first step in defining a consistent contraction of this algebra we write the $\mathfrak{sl}(2)$ algebra as $\mathfrak{so}(2,1)_k$,
\eq{
[\mathfrak{J}^a_n,\mathfrak{J}^b_m] = \epsilon^{ab}{}_c\,\mathfrak{J}^c_{n+m} - \frac{\hat{k}}{2} n\, \eta^{ab}\,  \delta_{n+m,\,0}
}{eq:carroll4}
where $\eta_{ab}$ (with $a,b = 0,1,2$) is the Minkowski metric with signature $(-1,1,1)$ and $\epsilon_{abc}$ is the three dimensional Levi-Civit\'a symbol with $\epsilon_{012} = +1$. The barred algebra is defined analogously, with the same value for the level $\hat k$.

We define the contracted generators in terms of the $\mathfrak{so}(2,1)_k$ current algebra generators $\mJ_n^a$, $\bar\mJ_n^a$ as follows ($i=1,2$)
\begin{subequations}
\label{eq:lalapetz2}
\begin{align}
P_n^i &= \frac{1}{\sqrt{\ell}}\,\big(\mJ_n^i + \bar\mJ_{-n}^i\big) \\ 
G_n^i &= \frac{1}{\sqrt{\ell}}\,\big(\mJ_n^i - \bar\mJ_{-n}^i\big) \\ 
J_n &= \mJ_{0,\,n} + \bar\mJ_{0,\,-n} \\ 
H_n &= \frac1\ell\,\big(\mJ_{0,\,n} - \bar\mJ_{0,\,-n}\big)\,.
\end{align}
\end{subequations}
The \.In\"on\"u--Wigner contraction $\ell\to\infty$ of the two $\mathfrak{so}(2,1)_k$ current algebras then yields the ASA
\begin{subequations}
\label{eq:carroll1}
\begin{align}
[J_n,\,P_m^i] &= \epsilon^i{}_j\, P^j_{n+m} \\
[J_n,\,G_m^i] &= \epsilon^i{}_j\, G^j_{n+m} \\
[P_n^i,\,G_m^j] &=-\epsilon^{ij}\,H_{n+m}  - kn\,\delta^{ij}\,\delta_{n+m,\,0}
\end{align}
\end{subequations}
where all commutators not displayed vanish, $k=\hat k/\ell$ and $\epsilon_{ij}=\epsilon_{0ij}$. Restricting the integers $n,m$ to zero  yields the (global) Carroll algebra
\begin{subequations}
\label{eq:carroll2}
\begin{align}
[J,\,P^i] &= \epsilon^i{}_j\, P^j \\
[J,\,G^i] &= \epsilon^i{}_j\, G^j \\
[P^i,\,G^j] &=-\epsilon^{ij}\,H
\end{align}
\end{subequations}
where we dropped the zero subscripts for notational simplicity. 

The flat space limit based on \eqref{eq:lalapetz2} is a peculiar one since it involves a square root of the AdS radius. In fact, the bulk counterpart of this contraction of the ASA is simultaneously a flat space limit and an ultra-relativistic limit.\footnote{
To avoid confusion let us note that the usual flat space limit also leads to an ultra-relativistic contraction of the asymptoic symmetry algebra in AdS$_3$, the two-dimensional conformal algebra. In our case the ultra-relativistic limit arises also in the bulk and leads to Carroll gravity, which is distinct from flat space Einstein gravity.
} This can be seen by noting that the Carroll algebra \eqref{eq:carroll2} also emerges when contracting the $\mathfrak{so}(2,1)\oplus\mathfrak{so}(2,1)$ Chern--Simons gauge algebra, so that the ensuing Chern--Simons theory is Carroll gravity, with the bilinear form, or ``Carroll trace'', defined by
\eq{
 \langle { \tt H},{\tt J}\rangle = -1 \qquad \langle { \tt P}_i,{\tt G}_j\rangle = \delta_{ij}\,.
}{eq:carroll6}

In \cite{Bergshoeff:2016soe} Brown--Henneaux-like boundary conditions were proposed for Carroll gravity. The boundary conditions induced by the contraction procedure above differ from these boundary conditions and yield a larger symmetry algbera, namely the loop algebra of the Carroll algebra. As we shall demonstrate below, non-chirally symmetric contractions are possible as well and lead to yet-another set of Carroll gravity boundary conditions. 

We do not discuss geometric aspects of the loop algebra boundary conditions, but we are going to perform such a discussion for the boundary conditions below, which are perhaps the simplest ones leading to a fluctuating Carroll boundary metric.

\subsubsection{Carroll gravity with fluctuating boundary metric}

The specific case we study here is taking the left moving sector to give an $\mathfrak{sl}(2)_k$ algebra (case 1) while the right moving sector is constrained to case 4 and hence gives a $\mathfrak{u}(1)_k$ current algebra. Again we switch to an $\mathfrak{so}(2,1)_k$ algebra. The full algebra before contraction then reads
\begin{subequations}
\begin{align}
    [\mathfrak{J}^a_n,\mathfrak{J}^b_m] & = \epsilon^{ab}{}_c\,\mathfrak{J}^c_{n+m} - \frac{\hat{k}}{2} n\, \eta^{ab}\,  \delta_{n+m,\,0} \\
    [\bar{\mathfrak{J}}_n,\bar{\mathfrak{J}}_m] & = \frac{\hat{k}}{2}\, n\, \delta_{n+m,\,0}
\end{align}
\end{subequations}
with the same definitions as below \eqref{eq:carroll4}.

We define new generators for $i = 1,2$ such that
\begin{equation}\label{limit_Carroll}
    G_n^{\,i} = \frac{2}{\sqrt{\ell}} \mathfrak{J}_n^i \,, \qquad
    K_n = \frac{1}{\ell}\left(\mathfrak{J}_{0,\,n} - \bar{\mathfrak{J}}_{-n} \right) \,, \qquad
    J_n = \mathfrak{J}_{0,\,n} + \bar{\mathfrak{J}}_{-n} \,.
\end{equation}
In the limit $\ell \to\infty$ the new generators satisfy the commutation relations
\begin{subequations}\label{ASA_Carroll}
\begin{align}
[K_n,J_m] & = k n \delta_{m+n,0} \\ 
[J_n,G^{\,i}_m] & = \epsilon^i{}_j G_{n+m}^j \\
[G_n^i, G_m^j] & = - 2 \epsilon^{ij} K_{n+m} - 2nk\,\delta^{ij} \delta_{n+m,0}\,.
\end{align}
\end{subequations}

As before, this \.In\"on\"u--Wigner contraction involves a square root of the AdS radius and is simultaneously a flat space limit and an ultra-relativistic limit. Here, this can be seen by taking a similar limit on the generators of the $SO(2,1)\times SO(2,1)$ isometry group of AdS$_3$. For the generators ${\tt J}^a, \bar{{\tt J}}^a$ with $a,b = 0,1,2$ such that
\begin{equation}
    [{\tt J}_a,{\tt J}_b] = \epsilon_{abc}{\tt J}^c \qquad 
    [\bar{\tt J}_a,\bar{\tt J}_b] = \epsilon_{abc}\bar{\tt J}^c
\end{equation}
we define a new set of generators as
\begin{equation}
    {\tt H} = \frac{1}{\ell}\left({\tt J}_0 - \bar{\tt J}_0 \right)\qquad 
    {\tt J} = {\tt J}_0 + \bar{\tt J}_0 \qquad
    {\tt G}_i = \frac{1}{\sqrt{\ell}}\left( {\tt J}_i - \bar{\tt J}_i \right) \qquad
    {\tt P}_i = \frac{1}{\sqrt{\ell}}\left( {\tt J}_i + \bar{\tt J}_i \right)\,.
\end{equation}
These generators span the Carroll algebra \eqref{eq:carroll2}.

In \cite{Bergshoeff:2016soe} a Chern--Simons theory for the Carroll algebra was discussed in the context of ultra-relativistic gravity. The infinite dimensional symmetry algebra \eqref{ASA_Carroll} can be seen as a particular extension of the Carroll algebra by imposing boundary conditions for Carroll Chern--Simons theory in the spirit of section 4 of \cite{Bergshoeff:2016soe}. Take the Chern--Simons connection $a_\vp$ to be
\begin{equation}\label{Carrollcon}
    a_{\vp} =  \mathcal{K}(t,\vp){\tt J} + \mathcal{J}(t,\vp) {\tt H} + \mathcal{J}^i(t,\vp) ({\tt G}_i  + {\tt P}_i)\,.
\end{equation}
For later purposes choose the $t$-component of the connection as
\eq{
a_t = \mu(t,\,\vp)\,{\tt H}\,.
}{eq:carroll7}
By the standard Chern--Simons calculation of the asymptotic charges
\begin{equation}
\delta Q[\lambda] = \frac{k}{2\pi} \oint \extd\vp \,\langle \lambda, \delta a_{\vp} \rangle
\end{equation}
together with the Carroll trace \eqref{eq:carroll6} these boundary conditions lead to asymptotic charges
\begin{equation}\label{Carcharge}
Q[\lambda] = \frac{k}{2\pi} \oint \extd\vp \left(- \lambda_{\tt J}\mathcal{J} - \lambda_{\tt H} \mathcal{K} + 2 \lambda^i_{{\tt G}} \mathcal{J}^j \delta_{ij} \right)\,.
\end{equation}
Here we parameterized the gauge parameter $\lambda$ as $\lambda = \lambda_{\tt J} {\tt J} + \lambda_{\tt H} {\tt H} + \lambda_{\tt G}^i ({\tt G}_i +{\tt P}_i)$. The state dependent functions transform under gauge transformations as
\begin{subequations}
\begin{align}
\delta \mathcal{K} & = \partial_{\vp} \lambda_{\tt J} \\
\delta \mathcal{J} & = \partial_{\vp} \lambda_{\tt H} + 2 \epsilon_{ij} \lambda_{\tt G}^j \mathcal{J}^j \\
\delta \mathcal{J}_i & = \partial_{\vp} \lambda_{\tt G}^i - 2 \epsilon_{ij} \lambda_{\tt G}^j \mathcal{K} + \epsilon_{ij} \lambda_{\tt J} \mathcal{J}^j\,. 
\end{align}
\end{subequations}
In terms of the Fourier modes 
\begin{subequations}
\begin{align}
K_n & = - \frac{k}{2\pi} \oint \extd \vp e^{in\vp} \mathcal{K} \\
J_n & = - \frac{k}{2\pi} \oint \extd \vp e^{in\vp} \mathcal{J}  \\
G_n^i & = \frac{k}{\pi} \oint \extd \vp e^{in\vp} \mathcal{J}^i 
\end{align}
\end{subequations}
the Poisson bracket algebra of the Carroll charges \eqref{Carcharge} leads to the ASA \eqref{ASA_Carroll}.

Thus, we have shown that starting from non-chiral boundary conditions in AdS$_3$ we can generate new boundary conditions for Carroll gravity through an \.In\"on\"u--Wigner contraction. We discuss geometric aspects of these boundary conditions, following the notation and choice of group element in \cite{Bergshoeff:2016soe}, $b=e^{\rho P_2}$. For the two-dimensional Carroll line-element we find
\eq{
\extd s^2_{(2)} = \big((\rho \mathcal{K}-\mathcal{J}^1)^2 + (\mathcal{J}^2)^2\big)\,\extd\vp^2 + 2\mathcal{J}^2\, \extd\vp\extd\rho + \extd\rho^2
}{eq:carroll8a}
and for the time-component we recover a similar result as in \cite{Bergshoeff:2016soe},
\eq{
\tau = \mu\,\extd t + \big(\mathcal{J} - \rho\mathcal{J}^1\big)\,\extd\vp\,.
}{eq:carroll9a}
For zero mode solutions (and going to a co-rotating frame where $\mathcal{J}^2 = 0$) the two-dimensional line-element \eqref{eq:carroll8a} represents a conical singularity centered at $\rho = \mathcal{J}^1/\mathcal{K}$.
The crucial difference to the boundary conditions in  \cite{Bergshoeff:2016soe} is that the two-dimensional metric \eqref{eq:carroll8a} is allowed to fluctuate to leading ($\rho^2$-)~order in the $\extd\vp^2$-component. Apart from this, the respective line-elements and time-components essentially match. 

A puzzling aspect of our new Carroll gravity boundary conditions above is that the solutions \eqref{eq:carroll8a}, \eqref{eq:carroll9a} (for simplicity assuming constant chemical potential $\mu$ and charge ${\cal K}$) appear to have an entropy, despite of the fact that there is no horizon anywhere. The entropy, which satisfies a first law by construction, follows from naively applying the general Chern--Simons result \cite{Bunster:2014mua}
\eq{
S = k\, \big|\langle a_t a_\vp \rangle\big| = k\,\big|\mu{\cal K}\big|\,.
}{eq:S}

\section{Discussion}\label{se:6}

In this final section we conclude with a couple of further checks and remarks on holographic aspects, as well as an outlook to generalizations and future developments.

In section \ref{se:6.1} we show that the canonical boundary charges are conserved. 
In section \ref{se:6.2} we demonstrate that we have a well-defined variational principle for the most general flat space boundary conditions.
In section \ref{se:6.3} we collect some miscellaneous remarks on flat space and Carroll holography, as well as on related recent developments.
In section \ref{se:6.4} we provide an outlook to some future research directions.

\subsection{Charge conservation}\label{se:6.1}

Using on-shell conditions the (retarded) time derivative of the canonical boundary charges \eqref{eq:lalapetz} can be brought into the form
\begin{equation}
    \partial_u Q[\epsilon] = \frac{k}{2\pi} \oint \extd \vp \left(\epsilon_M^0 \partial_\vp \mu_L^0 + \epsilon_L^0 \partial_\vp \mu_M^0 - 2\epsilon_M^+ \partial_\vp \mu_L^- - 2\epsilon_L^+ \partial_\vp \mu_M^- - 2\epsilon_M^- \partial_\vp \mu_L^+ - 2\epsilon_L^- \partial_\vp \mu_M^+   \right)
    \label{eq:con1}
\end{equation}
that makes charge conservation manifest for constant chemical potentials $\mu_L, \mu_M$, since in this case $\partial_u Q[\epsilon]=0$. Note that even when the chemical potentials are $\vp$-dependent, the time derivative of the charges \eqref{eq:con1} is state-independent, so that we always have the conservation equation $\partial_u \delta Q = 0$ for the variation of the charges.

\subsection{Variational principle}\label{se:6.2}

The variational principle in the Chern--Simons formulation is well-defined for our boundary conditions \eqref{eq:bms4}-\eqref{eq:bms7}, provided we add the same boundary term as in \cite{Grumiller:2016pqb},\footnote{In the special case of section \ref{se:5.3} the chemical potentials depend on the state-dependent functions and hence there these results do not apply. In fact, in that case no boundary term is needed for a well-defined variational principle \cite{Detournay:2016sfv}.}
\eq{
\Gamma_{\textrm{\tiny CS}}[A]=I_{\textrm{\tiny CS}}[A] - \frac{k}{4\pi}\,\int_{\partial\mathcal M}\!\!\!\extd u\extd\vp\,\langle A_u A_\vp\rangle\,.
}{eq:vp}
The first variation of the full action $\Gamma_{\textrm{\tiny CS}}$ then vanishes.
\eq{
\delta\Gamma_{\textrm{\tiny CS}}\big|_{\textrm{\tiny EOM}} = - \frac{k}{2\pi}\,\int_{\partial\mathcal M}\!\!\!\extd u\extd\vp\,\langle a_\vp \,\delta a_u\rangle = 0
}{eq:checkvp}
The expression on the right hand side of \eqref{eq:checkvp} has the familiar holographic form $\textrm{vev}\times\delta\,\textrm{source}$, with the six vevs $\cL^a$, $\cM^a$ contained in $a_\vp$ and the six sources $\mu_L^a$, $\mu_M^a$ contained in $a_u$. The results above are essentially the same as in AdS$_3$ \cite{Grumiller:2016pqb}. 

We have not investigated the second order analog of the results above, i.e., we do not know what type of boundary terms need to be added to the Einstein--Hilbert action \eqref{eq:bms1} in order to render the variational principle well-defined for arbitrary metric fluctuations that preserve the boundary conditions \eqref{genFG}-\eqref{eq:gtphi}. We leave this issue for future investigations and mention on the next two pages further loose ends.

\subsection{Remarks on flat space and Carroll holography and on recent developments}\label{se:6.3}

It would be interesting to study holographic implications of our new boundary conditions (see \cite{Bagchi:2016bcd} and references therein for some literature on flat space holography). In particular, it would be gratifying to provide a field theory interpretation. Our results actually suggest such an interpretation analogous to the AdS$_3$ case discussed in \cite{Grumiller:2016pqb}, namely as an $\mathfrak{isl}(2)_k$ WZW model, since our physical Hilbert space falls into representations of the affine $\mathfrak{isl}(2)_k$ algebra. The chemical potentials $\mu_{L,\,M}$ are then reinterpreted as sources for corresponding operators, whose vevs are determined by the state-dependent functions ${\cal L}, \,{\cal M}$. As mentioned above, the variation of the on-shell action \eqref{eq:checkvp} already has the form expected for a holographic 1-point function, but it would be good to push this correspondence further and check at least the 2- and 3-point functions.

Concerning Carroll holography, the results in this paper suggest to adopt a similar strategy as for some aspects of flat space holography, namely to start with well-established results in AdS$_3$/CFT$_2$ and to contract these results suitably. To give just one example, all correlation functions of the Galilean/Carrollian analog of the stress tensor can be obtained either by direct computation or by contraction from corresponding AdS/CFT results \cite{Bagchi:2015wna}. It seems plausible that similar manipulations will be useful for Carroll holography. Additionally, it could be useful to generalize the Grassmanian approach applicable to flat space contractions \cite{Krishnan:2013wta} to the Carrollian case.

There are numerous further aspects and recent developments related to our work, which we have not addressed. We mention now some of them. Recent extensions, elaborations and modifications of the original BMS symmetries were addressed in \cite{Compere:2016jwb, Compere:2016hzt, Penna:2017bdn, Barnich:2017ubf,Troessaert:2017jcm}, see also references therein. It was shown in \cite{Hartong:2015usd} in the context of flat space holography in three dimensions that null infinity can be interpreted as Carrollian geometry. It may be useful to understand if/how that construction relates to the new Carrollian limits studied in this paper. 

Finally, during the past year, partly inspired by the soft hair proposal \cite{Hawking:2016msc, Hawking:2016sgy}, a particular research focus was on near horizon aspects, specifically on new boundary conditions that allow a simple and natural interpretation in terms of an expansion around a non-extremal horizon. A zoo of different symmetry algebras was discovered in this way: a centerless warped conformal algebra \cite{Donnay:2015abr,  Donnay:2016iyk}, a twisted warped conformal algebra \cite{Afshar:2015wjm}, two $\mathfrak{u}(1)_k$ current algebras \cite{Afshar:2016wfy, Afshar:2016kjj, Afshar:2016uax} or BMS$_3$ \cite{Carlip:2017xne}.\footnote{%
It is not yet clear which of these approaches is the most fruitful one. A striking feature of the approach advocated by us is the universality of the entropy $S$ in terms of near horizon zero mode charges $J_0^\pm$, given by $S=2\pi(J_0^+ + J_0^-)$, which considerably generalizes the Bekenstein--Hawking or Wald entropy formulas. This is so, since our result for entropy applies not only to Einstein gravity in AdS$_3$ \cite{Afshar:2016wfy} or flat space \cite{Afshar:2016kjj}, but also to higher spin theories in AdS$_3$ \cite{Grumiller:2016kcp} or flat space \cite{Ammon:2017vwt} and higher derivative theories (with or without gravitational Chern--Simons term) \cite{Setare:2016vhy}.} 
It could be interesting to reconsider some of these proposals in the light of the new (and more general) asymptotic boundary conditions proposed in the present work.

\subsection{Outlook}\label{se:6.4}

While we have provided the most general set of flat space boundary conditions in three-dimensional Einstein gravity in the sense that we have the maximal number of conserved charges that is conceivable in this theory, infinitely many deformations of our boundary conditions are obtained by allowing for suitable relationships between chemical potentials $\mu_{L,\,M}$ and charges ${\cal L}, \,{\cal M}$,
\eq{
\mu_{L,\,M}^n = \mu_{L,\,M}^n({\cal L}^l,\,{\cal M}^m)\,,
}{eq:muQ}
where the functional dependence may involve also derivatives. The word ``suitable'' here means that the six relations \eqref{eq:muQ} should not spoil integrability of the canonical boundary charges. A specific set of such relations for AdS$_3$ Einstein gravity based on the Korteweg--de~Vries hierarchy was discovered in \cite{Perez:2016vqo} where the Brown--Henneaux boundary conditions were deformed. Their analysis generalizes to flat space \cite{ikdv}.\footnote{DG thanks Javier Matulich and Ricardo Troncoso for announcing these results in private discussions.} It would be interesting to apply similar deformations to our boundary conditions.

It could be rewarding to recast more of our results in the metric formulation, since this may help in lifting our boundary conditions to higher dimensions where no Chern--Simons formulation exists. Such generalizations are interesting in itself and have the potential for novel applications in holography. For more remarks on holographic aspects see sections \ref{se:6.2} and \ref{se:6.3} above.

Finally, in our discussion of new boundary conditions of Carroll gravity we have barely scratched the surface, and there are numerous open questions that call for further investigations. For instance,  which other boundary conditions can be obtained from similar contractions? Or how to explain the puzzling (though not necessarily trustworthy) result that the Carroll geometries \eqref{eq:carroll8a}-\eqref{eq:carroll9a} appear to have an entropy \eqref{eq:S}? 

In conclusion, we consider it gratifying that the asymptotically flat story initiated by Bondi, van~der~Burg, Metzner and Sachs in the 1960ies continues to inspire new and surprising developments more than half a century later, even in the apparently simple case of three-dimensional gravity.

\subsection*{Acknowledgments}

We are grateful to Geoffrey Comp\`ere for editing the {\em BMS Focus Issue} for Classical and Quantum Gravity and for inviting us to contribute to it. We thank Hern\'an Gonz\'alez for discussions.  DG thanks Glenn Barnich for popularizing the beautiful word ``Cardyology'' \cite{Barnich:2017ubf}. 
We thank the Galileo Galilei Institute for Theoretical Physics (GGI) for the hospitality during the completion of this work. In addition MR thanks the Istituto Nazionale di Fisica Nucleare (INFN) as well as the Italian Banking Foundation Association for partial support. DG and WM are grateful for the stimulating atmosphere of the
focus week {\it Recent developments in AdS$_3$ black-hole physics} within the program ``New Developments in AdS$_3$/CFT$_2$ Holography'' in April 2017 at GGI. 

This work was supported by the Austrian Science Fund (FWF), projects P~27182-N27 and P~28751-N27. 
The research of MR is supported by the ERC Starting Grant 335146 ``HoloBHC''.

\section*{References}

\providecommand{\href}[2]{#2}\begingroup\raggedright\endgroup


\end{document}